\documentclass[11pt,letterpaper]{article}

\usepackage{authblk}
\usepackage{xcolor}
\usepackage{array}
\usepackage[title]{appendix}

\title{\Large Contemplating electromagnetic phenomena in lived experience 
through somatic meditation}
\author[1]{\normalsize Zosia Krusberg}
\author[2]{Andrew Feldman}
\author[3]{Elam Coalson}
\affil[1]{Department of Physics, University of Chicago}
\affil[2]{Steinhardt School, New York University}
\affil[3]{Plastic and Reconstructive Surgery, University of Chicago Hospitals}

\date{}

\begin{document}
\maketitle



\section{Introduction}

One of the objectives of the undergraduate physics curriculum is for students to become aware of the connections between formal physical principles and personal experience \cite{NRC2013}. However, research has shown that awareness of connections between the abstract and the experiential tends to deteriorate, sometimes significantly, following instruction in undergraduate physics courses \cite{Redish1998,Adams2006}.  Although this phenomenon has been discussed extensively in the literature, few pedagogical interventions have been designed or implemented to address this particular weakness in undergraduate physics instruction \cite{Smith2006,Chiaverina2012,Riendeau2014,Beck2016}.

In this work, we show that a contemplative practice consisting of a somatic meditation and a contemplation deepens students' awareness of the connections between formal physical principles and personal experience by deliberately drawing their attention to electromagnetic phenomena in their surroundings. In this process, students also note interdisciplinary connections between electrodynamic principles and chemical and biological systems. We also find that the contemplative practice awakens an intrinsic motivation to understand electromagnetic theory, as well as an appreciation for the somatic, affective, and cognitive benefits of a contemplative practice.

\section{Method}
\label{method}

\subsection{Contemplative practices in higher education}

Contemplative practices, which hold a central place in many spiritual and philosophical traditions, are characterized by an introspective exploration of one's personal experience in the present moment. Such practices take a wide variety of forms, including, but certainly not limited to, meditation and prayer, ceremony and ritual, vigil and pilgrimage, yoga and t'ai chi ch'uan.  Over the last several decades, studies within the emerging field of contemplative sciences have shown that contemplative practices---meditation in particular---can have substantial benefits in secular contexts.  In clinical psychology and medicine, for instance, mindfulness-based stress reduction (MBSR) and mindfulness-based cognitive therapy (MBCT) have had positive psychological and physiological effects on patients suffering from anxiety, depression, post-traumatic stress disorder, and various other chronic conditions \cite{Kabat-Zinn2006,Bohlmeijer2010,Hayes2011,Buchholz2015}.  The benefits of meditation have also been explored extensively in contexts such as sports \cite{Rooks2017}, the workplace \cite{Slemp2019}, and the criminal justice system \cite{Ferszt2015,Yoon2017}.

Contemplative practices have also been integrated alongside traditional pedagogical activities in educational institutions.  In higher education, meditation has been shown to improve the social, emotional, and cognitive health of both students and teachers \cite{Shapiro2008,Morrison2014,Short2015}, and to support learning, creativity, and the discovery of personal meaning in course material \cite{Palmer2010,Zajonc2013}.  Contemplative practices have now been implemented in courses in the humanities, the social sciences, and the natural sciences \cite{Hart2004,Bush2011,Barbezat2014}, including biology \cite{Haskell2012,Haskell2017}, chemistry \cite{Francl2016}, earth science \cite{Schneiderman2013}, environmental science \cite{Wapner2016,Wamsler2018}, mathematics \cite{Wolcott2013,Morgan2018}, and physics \cite{Zajonc2010,Krusberg2018}.

\subsection{The contemplative practice}

The contemplative practice developed in this work consists of a guided somatic meditation\footnote{In somatic meditation, the body, rather than the mind, is the object of meditation.} and contemplation. In the somatic meditation, students are instructed to direct their attention into and throughout their bodies, gradually expanding their awareness into the surrounding space with the help of their senses. Students are then asked to contemplate manifestations of electromagnetic phenomena in their surroundings while maintaining a light somatic awareness. An abbreviated version of the contemplative practice guidelines is shown in Table 1\footnote{ The handout describing the practice to students can be accessed at http://bit.ly/cont-em}.

\begin{table}[h]
\begin{center}
\scriptsize
\begin{tabular}{| m{12cm} |}
\hline
{\it Meditation} \\
Begin by stretching out your arms and legs, wiggling your fingers and toes, and loosening up and relaxing your body. Find a comfortable seated position and become aware of your body. Bring your attention to your seat. Feel the weight of your body and the support of the Earth. Let your body settle. Close your eyes. Concentrate on your sense of hearing and note any sounds. Sense the space around you, expanding beyond your immediate surroundings. Bring your attention into the center of your chest, placing your hand gently over your heart. Note the quality of your presence. Let your awareness spread throughout your body. Then, open your eyes and extend your awareness into the space around you.\\
\\
{\it Contemplation} \\
Maintaining an awareness of your body and the space around you, contemplate the manifestations of electromagnetic phenomena in your surroundings. Some may be immediately apparent, some may be less so. Questions, distractions, and insights may arise. Make room for all of it. Whatever your experience may be, it is relevant and valuable. \\
\hline
\end{tabular}
\end{center}
\caption{Abbreviated contemplative practice guidelines.}
\end{table}

\subsection{Implementation}

The contemplative practice was integrated into three calculus-based courses in classical electrodynamics at Northwestern University. Two courses consisted primarily of prospective physics majors, and one of students majoring in the engineering sciences. In each course, the contemplative practice was assigned in the third week of a nine-week quarter. Following completion of the contemplative practice, students submitted a written reflection on their experience. These reflections were graded on completion only. A total of 66 students completed the practice, of whom 15 were prospective physics majors.  

\subsection{Analysis}

Student reflections were collected electronically, stripped of personally identifying information, and coded. The coding process consisted of two phases: in the first phase, five common themes were identified, and in the second phase, phrases associated with each theme were labeled and catalogued. In order to ensure consistency, coding of student reflections was completed by one of the authors, and reviewed by another. A summary of the results is shown in Table 2.

\section{Results}
\label{results}

\subsection{Awareness of electromagnetic phenomena}

In their reflections, 62 out of 66 students described electromagnetic phenomena brought to their attention by the contemplative practice. 

Students noted the relationship between the macroscopic forces analyzed in their classical mechanics courses and fundamental electromagnetic interactions, for instance, the relationship between the normal force between the chair and their bodies, and the frictional force between their feet and the floor, and electrostatic interactions between electrons in the two objects.

Students frequently mentioned their personal technologies (phones, tablets, laptops) and common household appliances (lighting, heating and cooling), recognizing that these tools depend on electrical circuits. Many students extended their awareness further, contemplating electrical grids, power plants, and sources of energy.

Light was another commonly discussed electromagnetic phenomenon. Students generally began by acknowledging that vision relies on the detection of electromagnetic waves. They noted the emission of light by appliances, electronics, and the Sun, as well as the scattering of light off objects in their surroundings. Many students discussed the many practical uses of electromagnetic radiation, including the emission and detection of radio waves for communication and x-rays for medical imaging. 

Students also reflected on the presence of electromagnetic phenomena on planetary and astronomical scales. They described the Rayleigh scattering of sunlight to produce the Earth’s blue sky and the colors of the sunset. They discussed electrical storms, planetary magnetic fields, aurorae, and the interaction between electromagnetic waves and planetary atmospheres. 

Following these descriptions, students often acknowledged that electromagnetic phenomena pervade their immediate experience. In one representative passage, a student wrote,

\begin{quote}
This contemplative practice helped me think about how little I understand and think about something that affects my life every single day. Without it my day to day life would be unrecognizable and it is hard to think of an aspect of my life that is completely independent of electromagnetism.
\end{quote}

Only four students did not mention electromagnetic phenomena in their reflections. Of those four, three discussed the somatic and mental benefits of their experience.

\subsection{Awareness of interdisciplinary connections}

In their reflections, 34 out of 66 students identified interdisciplinary connections between electrodynamics and biological and chemical systems.

Students wrote extensively about the application of electrodynamic principles to human physiology. They recognized the pervasive effects of electrical forces on their bodies, including the currents regulating heartbeats via pacemaker cells, the flow of ions through neurons, and the conversion of physical phenomena to electrical signals in the brain. A few students noted that selective ion channels and action potentials were the results of electromagnetic interactions. 

Students also drew connections between electromagnetic theory and their knowledge of chemistry and biochemistry. In particular, many noted the importance of electron flows in both physics and biochemistry,

\begin{quote}
I thought about how I learned in biochemistry class that almost everything in life can be reduced down and explained as a flow of electrons, meaning that electricity plays a role in every aspect of our lives, not just in electronics and technology.
\end{quote}

Finally, in discussions of electromagnetic radiation, students frequently mentioned the relationship between the energy levels of electrons in atoms and the colors of light emitted by different chemical elements.

\subsection{Curiosity}

Beyond acknowledging the prevalence of electromagnetic phenomena in their lived experience, 35 out of 66 students described experiencing an emergent curiosity about their observations, often following up descriptions of phenomena with thoughtful questions.

Students asked questions about the shocks they experience from the buildup of electric charge on their bodies and clothing. They wanted to understand the mechanisms underlying their personal technologies. Questions arose around the physical senses, including general questions about how the brain processes physical stimuli, how mechanical waves are converted into what the brain interprets as sound, and how the brain perceives color. Students also expressed interest in the physics underlying human physiology and medical devices such as EKGs, EEGs, and MRIs. 

Students also contemplated a number of big-picture questions about the fundamental principles of elementary particle physics. For instance, students expressed general curiosity about the properties of elementary particles and the nature of their interactions. Many wondered whether the universe could exist without electromagnetic interactions. One student pondered whether the four fundamental interactions would ever be unified in a single theory, and expressed a desire to learn more about theoretical high energy physics. 

Many students expressed a sense of intrinsic motivation to attain a deeper understanding of the principles of electrodynamics. This especially followed realizations that the electromagnetic phenomena they identified were generally more complex than the systems analyzed in class. One student wrote,

\begin{quote}
I also notice with dismay the utter lack of point charges and/or simple two charge electrostatic systems. On the other hand, this makes me both motivated and excited to dive deeper into electromagnetism: to eventually understand with more with more mathematical and physical precision the complex systems that surround me.
\end{quote}

\subsection{Appreciation}

In their reflections, 53 out of 66 students expressed some form of appreciation for the somatic, affective, and cognitive effects of the practice.

Students reported that they enjoyed the opportunity to experience a sense of embodiment and relaxation. One student described feeling like the practice allowed them to align their mind with their body. Students mentioned appreciating the effects of the practice on their affective state: they enjoyed a heightened sense of awareness of themselves, their surroundings, and their experience, as well as a feeling of inner peace. 

Finally, students experienced a number of effects on their cognitive function, including a sense of clarity and focus. They particularly appreciated the feeling of curiosity that emerged from the practice, as well as the sense of excitement to learn more about the principles of classical electrodynamics:

\begin{quote}
A contemplative practice would never have occurred to me as part of a physics course \ldots After experiencing this practice in meditation and contemplation, however, I now wonder why such exercises are not more commonly encouraged in science courses. By meditating and contemplating my personal connection with electromagnetic phenomena, I found both relaxation and focus, which then allowed me to find a curiosity and interest in physics which I had not previously felt.
\end{quote}

As a consequence of these positive experiences, many students expressed a desire to integrate a contemplative practice into their daily lives.

\subsection{Skepticism}

In their reflections, 9 out of 66 students described feeling some initial hesitation about the practice. In some cases, this took the form of an initial judgment about the value of a contemplative practice in a physics course. These students described feeling somewhat uneasy, apprehensive, and confused about the purpose of the assignment. In other cases, students experienced difficulty settling down, relaxing, and feeling present, especially in contrast with their typically active and productive mindset. One student described feeling a sense of discomfort with complete silence. However, every one of these students ultimately expressed an appreciation for the practice. In one typical response, a student said,

\begin{quote}
When I read the handout for this assignment, I was very confused. The concept of meditation for a physics class seemed quite strange and foreign. However, \ldots I really enjoyed this assignment. Now I am much more mindful of electromagnetic phenomena in daily life, and I think I am going to start meditating regularly.
\end{quote}

Only one student questioned whether the meditative element of the practice contributed to a deepened awareness of electromagnetic phenomena.

\begin{table}[h]
\begin{center}
\scriptsize
\begin{tabular}{| m{3cm} | m{2cm} | m{7cm} | }
\hline
{\bf Theme} & {\bf Prevalence} & {\bf Examples} \\
\hline
Awareness of electromagnetic phenomena & 94\% & Macroscopic forces (normal, friction, tension), personal technologies (phones, tablets, laptops), household appliances (lamps, refrigerators, heaters, air conditioners, microwaves, washers and dryers), electricity (electrical outlets, electrical networks, power plants, sources of energy), electromagnetic radiation (light, color, heat), atmospheric and astronomical phenomena (lightning, aurorae, interaction between electromagnetic waves and atmospheres, planetary magnetic fields, electromagnetic processes in stars and galaxies) \\
\hline
Awareness of interdisciplinary connections & 52\% & Human physiology (perception, central and peripheral nervous system, cardiovascular system, electrolyte replacement), medical imaging (x-rays, MRI, EEG, EKG, ECG), biology (bioelectrogenesis, electroreception),  biochemistry, chemistry (electrical bonds, electrical properties of water), psychology (effects of sensory deprivation), computer science (hardware engineering, artificial intelligence) \\
\hline
Curiosity & 53\% & Electrostatics (electric shocks), technologies and appliances (phone chargers, motion detection, transmission of musical signals, efficiency of lightbulbs, electromagnetic shielding), human physiology (perception, detection of other regions of electromagnetic spectrum, effect of electric and magnetic fields on human health), theoretical physics (existence of two electrical charges, relative strengths of four fundamental interactions, grand unified theories) \\
\hline
Appreciation & 80\% & Awareness (self, affect, cognition, surroundings), somatic effects (release of tension, relaxation, embodiment), affective effects (tranquility, joy, excitement), cognitive effects (clarity, focus, motivation) \\
\hline
Skepticism & 14\% & Somatic (physical discomfort, difficulty sitting still and in silence), cognitive (unease, apprehension, confusion) \\
\hline
\end{tabular}
\end{center}
\caption{Summary of results.}
\end{table}

\section{Discussion} 
\label{discussion}

In this work, we show that a contemplative practice consisting of a somatic meditation and contemplation draws students’ attention to a variety of electromagnetic phenomena in their surroundings, allowing them to deepen their awareness of the relevance of classical electrodynamics to their lived experience. Based on the students’ reflections, the somatic meditation contributed in two major ways. First, the meditation prompted students to take pause and reduce both inner and outer distractions, which allowed them to achieve a state of mental openness toward the contemplation that followed. Second, the students’ identification of electromagnetic phenomena closely followed the progression of the guided meditation: as students’ awareness entered their bodies and the space around them, they noted the applicability of electromagnetic theory to their physical senses, the physiology of the human body, the technologies that permeate their everyday lives, and the physical universe on the largest scale.

Students expressed particular interest in the human body and physiology, including perceptions, the nervous system, and the cardiovascular system. Studies have shown that students generally perceive physics to be detached from their studies in chemistry and biology, resulting in a fragmented view in which physics does not contribute to their understanding of chemical and biological systems \cite{Geller2019}. By failing to explore interdisciplinary connections between STEM fields that are meaningful to students, traditional physics curricula therefore miss an opportunity both to build disciplinary coherence and to motivate students majoring in other physical sciences or the life sciences, who constitute an overwhelming majority of students in introductory physics courses \cite{NRC2013}. By drawing students’ awareness into their bodies, our contemplative practice offers an opportunity for students to naturally and meaningfully relate electrodynamic principles to chemical and biological systems. 

Finally, the contemplative practice had a notably positive impact on the students’ somatic, affective, and cognitive states. Research has shown that contemplative practices, particularly those incorporating elements of traditional mindfulness meditation, can help support student mental health under academic stress by fostering important cognitive processes such as attention and information processing, as well as by decreasing stress and anxiety \cite{Goleman2017}. Although these effects would not persist following a single meditation session, the contemplative practice may serve to introduce students to the benefits of integrating a mindfulness meditation practice into their daily lives. Many students mentioned the contemplative practice specifically in their course evaluations as a particularly impactful element of the course. 

The authors would like to encourage their colleagues to consider implementing contemplative practices into their curricula. This paper describes only one possible manifestation of such a practice, and the pedagogical objectives it addresses. However, the possibilities are endless. In our experience, students were overwhelmingly openminded about engaging in a highly unconventional activity for a university physics course. Meanwhile, the practice did not contribute significantly to our grading burden since credit was awarded on completion. We would simply advise that our colleagues first engage in their contemplative practice themselves, in order to attain a general sense of students’ possible experience with the practice. We also recommend exploring the resources offered by the Association for Contemplative Mind in Higher Education\footnote{https://acmhe.org/}, which has aided us tremendously in the development and implementation of a growing number of contemplative practices.


\clearpage

\addcontentsline{toc}{section}{References}
\bibliographystyle{unsrt}
\bibliography{standard-bibliography}

\begin{thebibliography}{10}

\bibitem{NRC2013}
{National Research Council}.
\newblock {\em {Adapting to a Changing World: Challenges and Opportunities in
  Undergraduate Physics Education}}.
\newblock The National Academies Press, Washington, DC, 2013.

\bibitem{Redish1998}
Edward~F. Redish, Jeffery~M. Saul, and Richard~N. Steinberg.
\newblock {Student expectations in introductory physics}.
\newblock {\em American Journal of Physics}, 66(3):212--224, 1998.

\bibitem{Adams2006}
Wendy~K. Adams, Katherine~K. Perkins, Noah~S. Podolefsky, Michael Dubson,
  Noah~D. Finkelstein, and Carl~E. Wieman.
\newblock {New instrument for measuring student beliefs about physics and
  learning physics: The Colorado Learning Attitudes about Science Survey}.
\newblock {\em Physical Review Special Topics -- Physics Education Research},
  2(010101):1--14, 2006.

\bibitem{Smith2006}
Donald Smith.
\newblock {Mechanics in the real world}.
\newblock {\em The Physics Teacher}, 44(3):144--145, 2006.

\bibitem{Chiaverina2012}
Chris Chiaverina.
\newblock {Taking physics class into the world}.
\newblock {\em The Physics Teacher}, 50(9):572--573, 2012.

\bibitem{Riendeau2014}
Diane Riendeau.
\newblock {Using the real world to teach physics}.
\newblock {\em The Physics Teacher}, 52(2):125, 2014.

\bibitem{Beck2016}
Judith Beck and James Perkins.
\newblock {The "Finding Physics" project: Recognizing and exploring physics
  outside the classroom}.
\newblock {\em J. Phys. J. Phys}, 54(8):466--468, 2016.

\bibitem{Kabat-Zinn2006}
Jon Kabat-Zinn.
\newblock {Mindfulness-based interventions in context: Past, present, and
  future}.
\newblock {\em Clinical Psychology: Science and Practice}, 10(2):144--156,
  2006.

\bibitem{Bohlmeijer2010}
Ernst Bohlmeijer, Rilana Prenger, Erik Taal, and Pim Cuijpers.
\newblock {The effects of mindfulness-based stress reduction therapy on mental
  health of adults with a chronic medical disease: A meta-analysis}.
\newblock {\em Journal of Psychosomatic Research}, 68(6):539--544, 2010.

\bibitem{Hayes2011}
Steven~C. Hayes, Matthieu Villatte, Michael Levin, and Mikaela Hildebrandt.
\newblock {Open, aware, and active: Contextual approaches as an emerging trend
  in the behavioral and cognitive therapies}.
\newblock {\em Annual Review of Clinical Psychology}, 7(1):141--168, 2011.

\bibitem{Buchholz2015}
Laura Buchholz.
\newblock {Exploring the promise of mindfulness as medicine}.
\newblock {\em Journal of the American Medical Association},
  314(13):1327--1329, 2015.

\bibitem{Rooks2017}
Joshua~D. Rooks, Alexandra~B. Morrison, Merissa Goolsarran, Scott~L. Rogers,
  and Amishi~P. Jha.
\newblock {“We are talking about practice”: The influence of mindfulness
  vs. relaxation training on athletes' attention and well-being over
  high-demand intervals}.
\newblock {\em Journal of Cognitive Enhancement}, 1(2):141--153, 2017.

\bibitem{Slemp2019}
Gavin~R. Slemp, Hayley~K. Jach, Austin Chia, Daniel~J. Loton, and Margaret~L.
  Kern.
\newblock {Contemplative interventions and employee distress: A meta-analysis}.
\newblock {\em Stress and Health}, 35(3):227--255, 2019.

\bibitem{Ferszt2015}
Ginette~G. Ferszt, Robin~J. Miller, Joyce~E. Hickey, Fleet Maull, and Kate
  Crisp.
\newblock {The impact of a mindfulness based program on perceived stress,
  anxiety, depression and sleep of incarcerated women}.
\newblock {\em International Journal of Environmental Research and Public
  Health}, 12(9):11594--11607, 2015.

\bibitem{Yoon2017}
Isabel~A. Yoon, Karen Slade, and Seena Fazel.
\newblock {Outcomes of psychological therapies for prisoners with mental health
  problems: A systematic review and meta-analysis}.
\newblock {\em Journal of Consulting and Clinical Psychology}, 85(8):783--802,
  2017.

\bibitem{Shapiro2008}
Shauna~L. Shapiro, Kirk~Warren Brown, and John~A. Astin.
\newblock {Toward the integration of meditation into higher education: A review
  of research}.
\newblock {\em Teachers College Record}, 113(3):493--528, 2008.

\bibitem{Morrison2014}
Alexandra~B. Morrison, Merissa Goolsarran, Scott~L. Rogers, and Amishi~P. Jha.
\newblock {Taming a wandering attention: Short-form mindfulness training in
  student cohorts}.
\newblock {\em Frontiers in Human Neuroscience}, 7:1--12, 2014.

\bibitem{Short2015}
Megan~M. Short, Dwight Mazmanian, Lana~J. Ozen, and Michel B{\'{e}}dard.
\newblock {Four days of mindfulness meditation training for graduate students:
  A pilot study examining effects on mindfulness, self-regulation, and
  executive function}.
\newblock {\em The Journal of Contemplative Inquiry}, 2(1):37--48, 2015.

\bibitem{Palmer2010}
Parker~J. Palmer and Arthur Zajonc.
\newblock {\em {The Heart of Higher Education: A Call to Renewal}}.
\newblock Jossey-Bass, San Francisco, CA, 2010.

\bibitem{Zajonc2013}
Arthur Zajonc.
\newblock {Contemplative pedagogy: A quiet revolution in higher education}.
\newblock {\em New Directions for Teaching and Learning}, 134:83--94, 2013.

\bibitem{Hart2004}
Tobin Hart.
\newblock {Opening the contemplative mind in the classroom}.
\newblock {\em Journal of Transformative Education}, 2(1):28--46, 2004.

\bibitem{Bush2011}
Mirabai Bush.
\newblock {Mindfulness in higher education}.
\newblock {\em Contemporary Buddhism}, 12(1):183--197, 2011.

\bibitem{Barbezat2014}
Daniel~P. Barbezat and Mirabai Bush.
\newblock {\em {Contemplative Practices in Higher Education}}.
\newblock Jossey-Bass, San Francisco, CA, 2014.

\bibitem{Haskell2012}
David~George Haskell.
\newblock {\em {The Forest Unseen: A Year's Watch in Nature}}.
\newblock Penguin Books, New York, NY, 2012.

\bibitem{Haskell2017}
David~George Haskell.
\newblock {\em {The Songs of Trees: Stories from Nature's Great Connectors}}.
\newblock Viking, New York, NY, 2017.

\bibitem{Francl2016}
Michelle Francl.
\newblock {Practically impractical: Contemplative practices in science}.
\newblock {\em Journal of Contemplative Inquiry}, 3(1):21--34, 2016.

\bibitem{Schneiderman2013}
Jill Schneiderman.
\newblock {Ground truth: Investigations of Earth simultaneously spiritual and
  scientific}.
\newblock In Jing Lin, Rebecca Oxford, and Edward Brantmeier, editors, {\em
  Re-Envisioning Higher Education: Embodied Pathways to Wisdom and Social
  Transformation}. Information Age Publishing, Charlotte, NC, 2013.

\bibitem{Wapner2016}
Paul Wapner.
\newblock {Contemplative environmental studies: Pedagogy for self and planet}.
\newblock {\em Journal of Contemplative Inquiry}, 3(1):67--83, 2016.

\bibitem{Wamsler2018}
Christine Wamsler, Johannes Brossmann, Heidi Hendersson, Rakel Kristjansdottir,
  Colin McDonald, and Phil Scarampi.
\newblock {Mindfulness in sustainability science, practice, and teaching}.
\newblock {\em Sustainability Science}, 13(1):143--162, 2018.

\bibitem{Wolcott2013}
Frank~Lucas Wolcott.
\newblock {On contemplation in mathematics}.
\newblock {\em Journal of Humanistic Mathematics}, 3(1):74--95, 2013.

\bibitem{Morgan2018}
Patricia Morgan and Dor Abrahamson.
\newblock {Applying contemplative practices to the educational design of
  mathematics content: A report from a pioneering workshop}.
\newblock {\em The Journal of Contemplative Inquiry}, 5(1):107--119, 2018.

\bibitem{Zajonc2010}
Arthur Zajonc.
\newblock {\em {Meditation as Contemplative Inquiry}}.
\newblock Lindisfarne Books, Great Barrington, MA, 2010.

\bibitem{Krusberg2018}
Zosia Krusberg and Meredith Ward.
\newblock {Classical physics and human embodiment: The role of contemplative
  practice in integrating formal theory and personal experience in the
  undergraduate physics curriculum}.
\newblock {\em Journal of Contemplative Inquiry}, 5(1):87--106, 2018.

\bibitem{Geller2019}
Benjamin~D. Geller, Julia Gouvea, Benjamin~W. Dreyfus, Vashti Sawtelle, Chandra
  Turpen, and Edward~F. Redish.
\newblock {Bridging the gaps: How students seek disciplinary coherence in
  introductory physics for life science}.
\newblock {\em Physical Review Physics Education Research}, 15(2):20142, 2019.

\bibitem{Goleman2017}
Daniel Goleman and Richard~J. Davidson.
\newblock {\em {Altered Traits: Science Reveals How Meditation Changes Your
  Mind, Brain, and Body}}.
\newblock Penguin Random House, New York, NY, 2017.

\end{thebibliography}

\end{document}